\documentclass[aip,jcp,preprint]{revtex4-1}

\usepackage{bbm}
\usepackage{color}
\usepackage{amsmath}
\usepackage{amssymb}
\usepackage{graphicx}
\usepackage{bm}

\newcommand{\HN}[1]{{#1}}

\begin{document}

\title{Nonequilibrium Thermodynamics and Steady State Density Matrix 
for Quantum Open Systems}

\author{Herv\'e Ness}

\affiliation{Department of Physics, Faculty of Natural and Mathematical Sciences,
King's College London, Strand, London WC2R 2LS, UK}

\begin{abstract}
We consider the generic model of a finite-size quantum electron system 
connected to two (temperature and particle) reservoirs. The quantum open
system is driven out of equilibrium by the presence of both a temperature 
and a chemical potential differences between the two reservoirs.
The nonequilibrium (NE) thermodynamical properties of such a quantum open 
system are studied for the steady state regime.
In such a regime, the corresponding NE density matrix is built on the so-called 
generalised Gibbs ensembles. From different expressions of the NE 
density matrix, we can identify the terms related to the entropy production 
in the system. We show, for a simple model, that the entropy production rate is 
always a positive quantity.
Alternative expressions for the entropy production are also obtained from the 
Gibbs~-~von Neumann conventional formula and discussed in detail. 
Our results corroborate and expand earlier works found in the literature.
\end{abstract}

%%%%%%%%%%%%%%%%%%%%%%%%%%%%%%%%%%%%%%%%%%
\maketitle

%%%%%%%%%%%%%%%%%%%%%%%%%%%%%%%%%%%%%%%%%%
%% Only for the journal Gels: Please place the Experimental Section after the Conclusions

%%%%%%%%%%%%%%%%%%%%%%%%%%%%%%%%%%%%%%%%%%
\section{Introduction}
\label{sec:intro}

The understanding of irreversible phenomena is a long-standing problem in 
statistical mechanics.
Explanations of the fundamental laws of phenomenological nonequilibrium (NE) thermodynamics
have been given and applied to quantum open systems for several decades
\cite{Spohn:1978,Alicki:1979}. More recent discussions on the origin of thermodynamical
laws at the nanoscale can be found in, for example, \cite{Allahverdyan:2004}.
Originally the weak coupling limit of a finite-size central region interacting with
thermal and/or particle baths was first considered \cite{Spohn:1978,Alicki:1979,Davies:1978,Spohn:1978b,Kosloff:2013}.
Methods for dealing with the strong coupling limit have
been recently developed ~\cite{Campisi:2009,Deffner:2011,Ajisaka:2012,Ludovico:2014,Esposito:2015,Topp:2015,Bruch:2016,Solano:2016}.
One can study the NE thermodynamical properties and the entropy production
in such systems when an external driving force is applied to the central
region. The time dependence of the external force can be arbitrary or periodic
\cite{Ludovico:2014,Esposito:2015,Bruch:2016,Ludovico:2016}.
The long-time limit behaviour of the NE thermodynamics presents also very 
interesting properties \cite{Topp:2015,Bruch:2016,Solano:2016}.

Indeed, after some time much longer than some typical relaxation times of 
the finite system, 
a steady state can be obtained. Such a state arises from the balance between 
irreversible processes (fluxes of particle and/or energy) and the driving forces 
induced by the reservoirs. 
The NE steady state presents some analogy to its equilibrium counterpart in the 
sense that
an equilibrium state represents a stationary state of a closed system, while
the NE steady state is the time-invariant state of an open system.
The fact that the NE steady state can be seen as a pseudo-equilibrium state
is central to the construction of the corresponding generalised Gibbs ensembles 
\cite{McLennan:1959,Zubarev:1974,Zubarev:1994,Zubarev:1996,Zubarev:1997,Morozov:1998}
and to the calculation of the entropy, heat or work production under NE conditions.

In the present paper, we construct such a generalised Gibbs expression for 
the NE density matrix and apply it to the calculation of the entropy
production in the system under the presence of
both a temperature difference and a chemical potential difference between the
two reservoirs.
Our approach has no restriction for the nature of the coupling (strong or weak) 
to the reservoirs, nor for the presence (or absence) of interaction between
particle in the central region.

In Section \ref{sec:NEdensmat}, we provide different, but fully
compatible, expressions for the NE density matrix and show that the new terms 
in the NE density matrix (new in the sense that they do not appear in the
equilibrium grand-canonical density matrix) are associated with the entropy production under 
the NE conditions. The entropy production rate is shown to be related to the fluxes of 
particle and heat across the system (Section \ref{sec:entropy}). 
We provide in Section \ref{sec:example_DeltaS}
some numerical calculations of the entropy production rate for a model system of a single
electron resonance coupled to two Fermi reservoirs. A comparison to earlier
results \cite{Esposito:2015} is also given.
In Section \ref{sec:entropy_more}, we consider the NE entropy production in the entire 
system obtained from the Gibbs~-~von Neumann expression based on the NE density matrix. 
Explicit derivations are provided in some limiting cases and it is shown that NE entropy
is produced not only in the central region but also in the reservoirs.
For the single resonance model, we also calculate the NE Gibbs~-~von Neumann entropy 
in the central region and present the corresponding
results in Section \ref{sec:example_SGibbs}.
Our approach corroborate and extend earlier existing results. Furthermore it opens a new 
route to the calculations of the full NE response functions of the system, such as the 
NE charge susceptibility \cite{Ness:2012} or the NE specific heat of the central region. 
%

%%%%%%%%%%%%%%%%%%%%%%%%%%%%%%%%%%%%%%%%%%
\section{Non-equilibrium steady state}
\label{sec:NEdensmat}

\subsection{System and initial conditions}
\label{sec:setup}

We consider a finite-size central region $C$, connected two electrodes (left $L$ 
and right $R$) acting as thermal and particle reservoirs. These electrodes
are described within the thermodynamics limit.
Initially they are at their own equilibrium, characterized by two temperatures 
$T_L$ and $T_R$, and by two chemical potentials $\mu_L$ and $\mu_R$. 
Furthermore, we ignore the interaction between particles in the electrodes, although 
the central region $C$ may contain such kind of interaction.
We are interested in steady state regime, and therefore we take the initial state 
of the system to be in the far remote past. The system 
is then characterised by an Hamiltonian $H_0$.
After all parts of the system are ``connected'' together and after some time elapses, 
the full system is considered to reach a NE steady state. 
The system is then characterised by the total Hamiltonian $H = H_0 + W$.

The questions related to the possibility of reaching a NE steady-state
have been addressed in \cite{Ruelle:2000,Tasaki:2003,Frohlich:2003,Tasaki:2006,
Maes:2010,Tasaki:2011,Moldoveanu:2011,Cornean:2011,Cornean:2014}.
It is also been argued that a system will always reach a steady-state
if it is a (or if it is connected to another) system in the thermodynamic 
limit regardless the presence (or absence) of adiabatic switching of
the interactions \cite{Ojima:1989,Cornean:2008}.

In the present paper, we consider that the full system is described by the Hamiltonian
$H = H_0 + W$
where $H_0$ is the non-interacting Hamiltonian $H_0 = H_L + H_C + H_R$
built from the three independent regions $L,C,R$.
The interaction $W$ is decomposed into several parts
$W =  V_{LC} + V_{CL} + V_{RC} + V_{CR} + V_C^{\rm int}$
where the interaction between particles
in region $C$ is given by $V_C^{\rm int}$ and the coupling between the $C$ region and the
$\alpha=L,R$ reservoirs is given by $V_{\alpha C}+V_{C\alpha}$.
Without specifying explicitly the form of $H_0$,
there exist different important commutation relations, i.e.
\begin{equation}
\label{eq:def_commut}
\begin{split}
[H_\alpha, H_\beta] = 0, \quad
[H_\alpha, N_\beta] = 0, \quad
[N_\alpha, N_\beta] = 0
\end{split}
\end{equation}
with $\alpha,\beta=L,C,R$ and $N_\beta$ is the occupation number operator
of the different regions $\alpha,\beta=L,C,R$.

Initially,  all regions $L,C,R$ are isolated and
characterised by their respective density matrix $\rho_\alpha$ with $\alpha=L,C,R$.
The macroscopic $L$ and $R$ regions are represented by a density matrix $\rho_{L,R}$ 
expressed in the grand canonical ensemble, 
with temperature $T_\alpha = 1/k\beta_\alpha$ and chemical potential $\mu_\alpha$
($\alpha=L,R$):

\begin{equation}
\label{eq:rho_LR}
\begin{split}
\rho_\alpha = \frac{1}{Z_\alpha} e^{-\beta_\alpha(H_\alpha-\mu_\alpha N_\alpha)} ,
\end{split}
\end{equation}
with $Z_\alpha={\rm Tr}_{(\alpha)}[e^{-\beta_\alpha(H_\alpha-\mu_\alpha N_\alpha)}]$
and ${\rm Tr}_{(\alpha)}$ implies a summation only over the states of the region $\alpha$.
The initial density matrix of the central region is assumed to take any arbitrary form
$\rho_C$ as this region is not in the thermodynamic limit.
Furthermore, considering $\rho_C$ to be given by a canonical or a grand 
canonical ensemble would imply the presence of the third reservoir,
which is not ideal in the present case. 
Therefore, we define $\rho_C$ from a microcanonical ensemble 
where
\begin{equation}
\label{eq:rho_C}
\begin{split}
\rho_C = \rho_C(H_C) = Z_C^{-1} \sum_n \vert C_n\rangle \delta_\Delta(\epsilon_n-E_C) \langle C_n\vert
\end{split}
\end{equation}
with the eigenstates $H_C\vert C_n\rangle = \epsilon_n\vert C_n\rangle$.
The $\delta_\Delta$ function is the ``regularized'' delta function defined by
$\delta_\Delta(\epsilon_n-E_C) = 1$ for $E_C\le \epsilon_n \le E_C + \Delta$ and 0
otherwise, and $Z_C={\rm Tr}_{(C)}[\delta_\Delta(H_C-E_C)]$.

The total density matrix $\rho_0$ as the non-interacting state defined by $H_0$
is given by the direct product $\rho_0 = \rho_L \otimes \rho_C \otimes \rho_R$.

\subsection{The NE density matrix $\rho^{\rm NE}$}
\label{sec:NErho}

In Ref.~\cite{Ness:2014c},
we used some concepts developed for asymptotic steady-state operators 
\cite{Doyon:2006,Fujii:2007,Gelin:2009,Hershfield:1993,Hyldgaard:2012,Bernard:2013}
and we have shown that the average of any arbitrary operator $A$ in 
the NE asymptotic steady state is given by
\begin{equation}
\label{eq:NEaverage}
\begin{split}
\langle A \rangle^{\rm NE} = {\rm Tr}[\rho^{\rm NE} A],
\end{split}
\end{equation}
where $\rho^{\rm NE}$ is the NE steady state density matrix.
One should also note that the trace in Eq.~(\ref{eq:NEaverage}) 
runs over all the states of the three $L,C,R$ regions.
The NE density matrix is defined from
$
\rho^{\rm NE}  = \Omega^{(+)} \rho_0 \Omega^{(+)-1}
$
where the Moeller operator \cite{GellMann:1953,Akhiezer:1981,Bohm:1993,Baute:2001},
characterising the asymptotic steady state, is given by:
$\Omega^{(+)} = {\rm lim}_{\tau\rightarrow -\infty}\ e^{i H \tau} e^{-i H_0 \tau}$.
Such an operator presents a central property, the intertwining relation
\cite{GellMann:1953,Akhiezer:1981,Bohm:1993,Baute:2001,Bernard:2013},
$\Omega^{(+)}H_0=H\Omega^{(+)}$, or equivalently
$H_0^{(+)} = \Omega^{(+)}H_0\Omega^{(+)-1} = H$.

By defining any asymptotic operator as  
$X_\alpha^{(+)} = \Omega^{(+)} X_\alpha \Omega^{(+)-1}$ ,
it can be shown from Eq.~(\ref{eq:def_commut}) that, 
when $X_\alpha = H_\alpha$ or $X_\alpha = N_\alpha$,
we have the following relation:
\begin{equation}
\label{eq:commut_X+}
\begin{split}
[ X_\alpha^{(+)} , H ] = \Omega^{(+)} [ X_\alpha , H_0 ] \Omega^{(+)-1} = 0 \ .
\end{split}
\end{equation}
Hence, any linear combination $Y^a = \sum_\alpha a_\alpha X_\alpha^{(+)}$
also commutes with $H$: $[Y^a,H]=0$. The quantity $Y^a$ will be called
a conserved quantity in the following.
Furthermore, for $Y^b = \sum_\beta b_\beta X_\beta^{(+)}$, it can be shown
that $[Y^a,Y^b]=0$
when $X_{\alpha,\beta} = H_{\alpha,\beta}$ or $N_{\alpha,\beta}$.
This follows from Eq.~(\ref{eq:def_commut}) and consequently from 
$[X_\alpha^{(+)}, X_\beta^{(+)}] = 0$.

We have now all the ingredients to study different expressions of the 
NE density matrix $\rho^{\rm NE}$ in the steady
state \cite{Fujii:2007,Bernard:2013}.
The latter can be recast as 
\begin{equation}
\label{eq:rho_NE}
\begin{split}
\rho^{\rm NE} & = \Omega^{(+)} \rho_0 \Omega^{(+)-1} 
 	        = \Omega^{(+)} \rho_L \Omega^{(+)-1} \Omega^{(+)} \rho_C \Omega^{(+)-1} \Omega^{(+)} \rho_R \Omega^{(+)-1} \\
& = Z_L^{-1} Z_R^{-1} e^{-\beta_L ( H_L^{(+)} - \mu_L N_L^{(+)} ) } \rho_C(H_C^{(+)})
    \ e^{-\beta_R ( H_R^{(+)} - \mu_R N_R^{(+)} ) } \\
& = Z^{-1}\ {\rm exp} \left\{
-\beta_L ( H_L^{(+)} - \mu_L N_L^{(+)} ) -\beta_R ( H_R^{(+)} - \mu_R N_R^{(+)} ) \right\} \
    \ \rho_C(H_C^{(+)}) ,
\end{split}
\end{equation}
where, in the last equality, we used $[X_\alpha^{(+)}, X_\beta^{(+)}] = 0$ and $Z=Z_L Z_R$.
Finally, one should also note that
$H_L^{(+)} + H_R^{(+)} = \Omega^{(+)} ( H_L + H_R ) \Omega^{(+)-1}
= \Omega^{(+)} ( H_0 - H_C ) \Omega^{(+)-1} = H - H_C^{(+)}$.

\subsection{Three equivalent expressions for  $\rho^{\rm NE}$}
\label{sec:3_NEdensmat}

Upon regrouping the different terms in the exponential of Eq.~(\ref{eq:rho_NE}), one obtains different,
but equivalent, expression for the density matrix.
The first expression is a generalisation of the density matrix derived by Hershfield 
\cite{Hershfield:1993,Hyldgaard:2012},
the third expression is the so-called McLennan-Zubarev NE statistical operator, while the
second expression is an intermediate between the two.

First, we have generalised in \cite{Ness:2014c} the results of 
Hershfield \cite{Hershfield:1993,Hyldgaard:2012}
to the presence of both a temperature and a chemical potential differences ($\beta_L\ne\beta_R$, $\mu_L\ne\mu_R$)
between the reservoirs.
The NE density matrix is then recast as follows:
follows
\begin{equation}
\label{eq:rho_NE_genHershfield}
\begin{split}
\rho^{\rm NE} 	= Z^{-1}\ {\rm exp} \left\{-\bar\beta ( H - Y^Q + Y^E ) \right\} \ 
                  \rho_C(H_C^{(+)}) \
		  e^{+\bar\beta H_C^{(+)} } ,
\end{split}
\end{equation}
where we have used the definitions and commutators given in Sec.~\ref{sec:NEdensmat}.
Note that the generalised Gibbs form of the NE density matrix in 
Eq.~(\ref{eq:rho_NE_genHershfield}) 
is given with an effective temperature $T_{\rm eff}$ defined 
from $\bar\beta=\frac{1}{2}(\beta_L+\beta_R)$. 
This temperature is different from the temperature of the reservoirs $T_{L,R}$ since 
$T_{\rm eff} = 1/k_B\bar\beta = 2 T_L T_R / (T_L + T_R)$.

The conserved quantities $Y^Q$ and $Y^E$ are related to the charge and energy currents
respectively via:
\begin{equation}
\label{eq:YQE}
\begin{split}
Y^Q =(\beta_L\mu_L N_L^{(+)} + \beta_R\mu_R N_R^{(+)}) / \bar\beta , \quad {\rm and} \quad
Y^E =(\beta_L-\beta_R)\frac{1}{2}(H_L^{(+)} - H_R^{(+)}) / \bar\beta.
\end{split}
\end{equation}

Second, following \cite{Bernard:2013}, one can re-expressed the density matrix
in a slightly different form (closer to the grand-canonical ensemble) involving 
quantities with a more explicit physical meaning.
Indeed, by writing 
$E^{(+)} = \Omega^{(+)} E \Omega^{(+)-1}$ with $E = \frac{1}{2}(H_L - H_R)$
and
\begin{equation}
\label{eq:Y_Q_Doyon}
\begin{split}
Y^Q & = \bar\mu N + \Delta_\mu Q^{(+)} / \bar\beta
\end{split}
\end{equation}
with $Q = \frac{1}{2}(N_L - N_R)$,
the NE density matrix takes the following form (with $X_{L+R}=X_L+X_R$):
\begin{equation}
\label{eq:rho_NE_2}
\begin{split}
\rho^{\rm NE} & = Z^{-1}\ {\rm exp} \left\{-\bar\beta ( H_{L+R}^{(+)} - \bar\mu N_{L+R}^{(+)} ) 
		+ \Delta_\mu Q^{(+)} - (\beta_L-\beta_R) E^{(+)} \right\} \
    \ \rho_C(H_C^{(+)}) \\
 & = Z^{-1}\ {\rm exp} \left\{-\bar\beta ( H - \bar\mu N ) 
	+ \Delta_\mu Q^{(+)} - (\beta_L-\beta_R) E^{(+)} \right\}
    \ \rho_C(H_C^{(+)}) \ e^{+\bar\beta ( H_C^{(+)} - \bar\mu N_C^{(+)} ) } \ ,
\end{split}
\end{equation}
where 
$\bar\mu  = ( \beta_L\mu_L + \beta_R\mu_R ) / ( \beta_L + \beta_R )$, 
$\Delta_\mu = \beta_L\mu_L - \beta_R\mu_R$ \cite{Note:ablinearcombination}.
Note that the passage from the first to the second line requires the use of an intertwining 
relation for $N$ \cite{note:N+}.

Furthermore, the initial density matrix $\rho_C$ could be given by any other form
different from Eq.~(\ref{eq:rho_C}) as such a choice is completely arbitrary. 
Indeed, in the steady state, the initial correlation vanishes \cite{Velicky:2010} 
and the final stationary properties should not dependent on the initial conditions
taken for the statistics of the central region. 
Hence, for convenience and to simplify the notation, we chose $\rho_C$ such that 
$\rho_C \ e^{+\bar\beta ( ...  ) } = \mathbbm{1}$
in Eq.~(\ref{eq:rho_NE}) and  Eq.~(\ref{eq:rho_NE_2}) \cite{Note:comment_on_rhoC}.

It is also important to note that, in Eq.~(\ref{eq:rho_NE_2}), 
the quantities $Q^{(+)}$ and $E^{(+)}$ are conserved quantities
and are directly related to the charge and energy currents.
Indeed, in the Heisenberg representation, the energy current operator is given
by $j_E(t)=\partial_t E(t) = \frac{i}{\hbar}[H,E(t)]$ and
the charge current operator $j_Q(t)$ is given by
$j_Q(t)=e\partial_t Q(t) = \frac{ie}{\hbar}[H,Q(t)]$.
This results permits us to connect the expressions Eq.~(\ref{eq:rho_NE}) and  Eq.~(\ref{eq:rho_NE_2})
for the NE density matrix to the third formulation, i.e. the McLennan~-~Zubarev NE statistical operator.

Such a statistical operator is given by \cite{Zubarev:1994,Tasaki:2006,Ness:2013}:
\begin{equation}
\label{eq:rhoNESOM}
\begin{split}
\rho^{\rm NESO} = \frac{1}{Z}\
{\rm exp} \left\{ - \sum_\alpha \beta_\alpha \left( H_\alpha - \mu_\alpha N_\alpha \right)  
+ \int_{-\infty}^0 {\rm d}u\ e^{\eta u} J_S(u) \right\} .
\end{split}
\end{equation}
The quantity $J_S(s)$ is called the non-systematic energy flows \cite{Tasaki:2006}
and is related to the entropy production rate of the system \cite{Frohlich:2003,Tasaki:2006,Ness:2013}.
It is given by 
\begin{equation}
\label{eq:Zub}
\begin{split}
J_S(u) = \sum_\alpha \beta_\alpha J_{S,\alpha}(u) \quad {\rm where} \quad
J_{S,\alpha}(u) =  \frac{d}{du} ( H_\alpha(u) - \mu_\alpha N_\alpha(u) )  
\end{split}
\end{equation}
where all operators are given in the Heisenberg representation, $A(u)=e^{iHu/\hbar} A e^{-iHu/\hbar}$.
In the literature, it is also customary to call $J_{S,\alpha}$
the heat current with results from the energy flux $J^E_\alpha(t)= \frac{d}{dt} H_\alpha(t)$ 
measured with respect to the so-called convective term 
$\mu_\alpha J^Q_\alpha(t) = \mu_\alpha\frac{d}{dt} N_\alpha(s) $
\cite{Sierra:2015}.
Because $J_S(u)$ is the sum of heat flows divided by the subsystem temperature, 
it is the entropy production rate of the whole system \cite{Frohlich:2003,Tasaki:2006}.
The time integration of $J_S(u)$ in Eq.~(\ref{eq:rhoNESOM}) provides the asymptotic steady state
value of the energy and charge fluxes $J^E_\alpha$ and $J^Q_\alpha$ respectively.
Hence the quantity $\int {\rm d}u J_S(u)$ is the entropy production in the 
NE steady state.

Recently we have shown~\cite{Ness:2014c} the full equivalence between the McLennan~-~Zubarev 
NE statistical operator $\rho^{\rm NESO}$ and the other expressions Eq.~(\ref{eq:rho_NE_genHershfield})
and Eq.~(\ref{eq:rho_NE_2}) for $\rho^{\rm NE}$.
The equivalence is based on the so-called Peletminskii lemma \cite{Peletminskii:1972} which states that the time
integral of an operator given in the Heisenberg representation (for example $\int {\rm d}u J_S(u)$) can be obtained
from an infinite series expansion of the time integral of the related quantities expressed in the interaction 
representation (see Appendix B in \cite{Ness:2014c}).

Hence the such an equivalence implies that the quantities $Y^{Q,E}$ in Eq.~(\ref{eq:rho_NE_genHershfield}) 
and the quantities $Q^{(+)},E^{(+)}$ in Eq.~(\ref{eq:rho_NE_2}) 
can be calculated from the same formal iterative scheme:
\begin{equation}
\label{eq:iteration_QE}
\begin{split}
Y = \sum_{n=0}^\infty Y_{n,I}  \quad {\rm and} \quad
\partial_t Y_{n+1,I}(t) = -\frac{i}{\hbar} [ {W}_I(t), Y_{n,I}(t) ] \ ,
\end{split}
\end{equation}
where we have used the notation $Y\equiv Y^Q, Y^E, Q^{(+)}$ or $E^{(+)}$ and 
the interaction representation, $A_I(t)=e^{iH_0t/\hbar} A e^{-iH_0t/\hbar}$, for all quantities.
The first values ($n=0$) of the series are
$Y_{0,I} = a^Q_L N_L + a^Q_R N_R$ when $Y \equiv Y^Q$ or $Q^{(+)}$
and
$Y_{0,I} = a^E (H_L - H_R )$ when $Y \equiv Y^E$ or $E^{(+)}$.
The different constants $a^Q_\alpha$ and $a^E$ are given by $a^Q_\alpha= { 2 \beta_\alpha\mu_\alpha }/( \beta_L+\beta_R )$ 
for  $Y=Y^Q$. For $Y=Q^{(+)}$ we have $a^Q_L = \frac{1}{2} = - a^Q_R$. For the energy flux,
we have $a^E = { \beta_L-\beta_R }/( \beta_L+\beta_R )$ for  $Y=Y^E$
or $a^E = \frac{1}{2}$ for $Y=E^{(+)}$.

\section{Entropy production}
\label{sec:entropy}

The equations Eq.~(\ref{eq:rho_NE_genHershfield}) and Eq.~(\ref{eq:rho_NE_2}) correspond to 
the most general expressions of the steady-state NE density matrix in the presence of both 
heat and charge currents (for a two-reservoir device).
We now use them to calculate the entropy production in the system under general NE conditions.

\subsection{Entropy production rate}
\label{sec:entropy_prod_rate}

As mentioned in the previous section, the different quantities $Y=Y^Q, Y^E, Q^{(+)}$ or $E^{(+)}$
are related to the entropy production (rate) in the system.
We can then define the NE entropy production $\Delta S^{\rm NE}$ in the steady state 
from Eq.~(\ref{eq:rho_NE_genHershfield}) and Eq.~(\ref{eq:rho_NE_2}) in the following way:
\begin{equation}
\label{eq:NESentropy}
\begin{split}
\Delta S^{\rm NE} / k_B
= \Delta_\mu \langle Q^{(+)} \rangle - (\beta_L-\beta_R) \langle E^{(+)} \rangle
= \bar\beta ( \langle Y^Q \rangle + \langle Y^E \rangle - \bar\mu \langle N \rangle)
= \int {\rm d}\tau\ \Delta \dot{S}^{\rm NE}(\tau) / k_B \ ,
\end{split}
\end{equation}
where $\Delta \dot{S}^{\rm NE}$ is the NE entropy production rate.
Hence, from the definition of $Q^{(+)}$ and $E^{(+)}$,
the NE entropy production rate is directly related to the asymptotic steady state
NE current of charge $\langle j_Q \rangle$ and energy $\langle j_E \rangle$:
\begin{equation}
\label{eq:NESentropy_rate}
\Delta \dot{S}^{\rm NE} / k_B 
= (\beta_L\mu_L - \beta_R\mu_R) \langle j_Q \rangle / e - (\beta_L-\beta_R) \langle j_E \rangle .
\end{equation}
\HN{Such a result has also been used in early work \cite{Note:comments_on_entropy}.}

A few remarks are now in order.
At equilibrium when $T_L=T_R$ and $\mu_L=\mu_R$, there are obviously no current
flows and no extra entropy is produced (apart from the equilibrium entropy arising from
the thermal fluctuations in the $L$ and $R$ reservoirs).
When $T_L=T_R$, there is no energy flow and there is a charge current when $\mu_L\ne\mu_R$. 
By convention, the NE averaged charge current $\langle j_Q \rangle$ is taken positive (negative)
when flowing from $L(R)$ to $R(L)$, i.e. when $\mu_L>\mu_R$ ($\mu_L<\mu_R$). 
Hence the contribution $(\mu_L - \mu_R) \langle j_Q \rangle$ to the entropy production is
always positive. 
Similarly, the NE averaged energy current $\langle j_E \rangle$ is taken positive (negative) 
when flowing from $L(R)$ to $R(L)$, i.e. when $T_L>T_R$ ($T_L<T_R$). 
Hence the contribution $-(\beta_L-\beta_R) \langle j_E \rangle$ to the entropy production is
also always positive.
However, the sign of the total contribution from $\langle j_Q \rangle$ and $\langle j_E \rangle$
to the entropy production (under general NE conditions) is not obvious without further investigation.
A general argument for the positiveness of the NE entropy production rate was given in \cite{Topp:2015}
and a few numerical examples were given in \cite{Esposito:2015}.

Using the same model system, i.e. the non-interacting single level coupled to two reservoirs, 
we provide in the next Section results for the entropy production rate for a wide range of parameters.

\subsection{An example}
\label{sec:example_DeltaS}

In the absence of interaction, the Hamiltonian for the central region $C$ is simply 
given by $  H^0_C   = \varepsilon_0 d^\dag d $
where $d^\dagger$ ($d$) creates (annihilates) an
electron in the level $\varepsilon_0$.
The non-interacting reservoirs are also described by a quadratic Hamiltonian 
$\alpha=L,R$ with $H_\alpha   = \sum_{i} \varepsilon_\alpha c^\dag_{\alpha i} c_{\alpha i} 
+ t_\alpha (c^\dag_{\alpha i} c_{\alpha i+1} + c.c.)$
where ${\alpha i}$ is an appropriate composite index to label the free electrons on the site $i$ 
of the $\alpha$ reservoirs.
The coupling between the central region and the electrodes is given via some hopping matrix 
elements $v_{\alpha}$, and we have 
$\sum_\alpha ( V_{C\alpha} + V_{\alpha C} ) = 
\sum_\alpha v_\alpha \left( c^\dag_{\alpha 0} d + d^\dag c_{\alpha 0}  \right)$. 
We recall that, by definition, we have $W = H^0_C + \sum_\alpha ( V_{C\alpha} + V_{\alpha C} )$. 
The only non vanishing anti-commutators are $\{d,d^\dag\}=1$ and  
$\{c_{\alpha i} , c^\dag_{\beta j}\} = \delta_{ij} \delta_{\alpha\beta}$.

The charge and energy currents can be calculated from the NE average expression in 
Eq.(\ref{eq:NEaverage}) \cite{Ness:2014c}, from asymptotic steady state scattering techniques 
\cite{Han:2006,Han:2007,Han:2007b,Han:2010,Han:2010b,Topp:2015} or 
from a NE Green's functions (NEGF) approach \cite{Esposito:2015,Bruch:2016}.
The full equivalence between the asymptotic steady state 
scattering and the NEGF techniques has been shown in \cite{Han:2012}. 

In \cite{Ness:2013}, we have stressed that calculating the NE averages
with the NE density matrix and 
the series expansion of the operators $Y$ in Eq.~(\ref{eq:iteration_QE}),
is equivalent to the NEGF approach in the steady-state regime. 
The Green's functions
are correlation functions whose thermodynamical averages
are formally identical to those given in Eq.(\ref{eq:NEaverage}).
Both perturbation series used in the NEGF approach and
in the derivations of the equations for the $Y$ operators 
start from the same nonequilibrium series expansion. 
They are just two different ways of summing that
series. For a non-interacting problem for which the series can
be resumed exactly, the NEGF and the NE density matrix with the $Y$ 
operators approach provide the same result \cite{Schiller:1995,Schiller:1998}. 
For an interacting system, one must resort to approximations to 
partially resum the series, and therefore the two approaches are 
similar only when the same approximations are used.
For the purpose of the present section, 
we then use the NEGF approach as the calculations are more straight forward in 
the non-interacting case. 
We also note that the NEGF formalism  permits us to include local interaction 
in the central region in a compact and self-consistent scheme, as we have done
in 
\cite{Dash:2010,Ness:2010,Dash:2011,Ness:2011,Dash:2012,Ness:2012,Ness:2014a,Ness:2014b}.
 
For the non-interacting system, the charge and energy currents are related to the 
transmission coefficient $T(\omega)$ of the junction via the moments $M_n$:
\begin{equation}
\label{eq:Landauer_currents}
\begin{split}
M_n = \frac{1}{\hbar} \int \frac{{\rm d}\omega}{2\pi}\ \omega^n\ T(\omega) (f_L(\omega)-f_R(\omega))
\ ,
\end{split}
\end{equation}
where $f_\alpha(\omega)$ is the equilibrium Fermi distribution of the reservoir $\alpha$.
The charge current is $\langle j_Q \rangle = e M_0$ and the energy current is
$\langle j_E \rangle = M_1$.
The transmission is obtained from 
$T(\omega) = G^r(\omega) \Gamma_L(\omega) G^a(\omega) \Gamma_R(\omega)$ where the NEGF $G^{r,a}$ 
are
given by $G^r(\omega) = [ \omega - \epsilon - \Sigma^r_{L+R}(\omega) ]^{-1}=(G^a(\omega))^*$, with
$\Sigma^r_{L+R}=\Sigma^r_L + \Sigma^r_R$ being the reservoirs self-energy. 
Furthermore we have
$\Gamma_\alpha(\omega)= \Sigma^a_\alpha(\omega) - \Sigma^r_\alpha(\omega)$ 
and the reservoir $\alpha$ self-energy is defined by
$\Sigma^r_\alpha(\omega)= v_\alpha^2 e^{-ik_\alpha(\omega)} / t_\alpha$ with 
the dispersion relation $\omega = \varepsilon_\alpha - 2 t_\alpha \cos(k_\alpha(\omega))$.

Figure 1 shows the NE entropy production rate $\Delta \dot{S}^{\rm NE}$ 
calculated for different transport regimes. The main conclusion is that $\Delta \dot{S}^{\rm NE}$
is always a positive quantity, as expected.
Such a behaviour is obtained for a system with a single chemical potential (see panel (a)
in Fig. 1). It is also obtained when there are both a chemical potential and temperature 
differences between the reservoirs, regardless of the respective direction of the charge 
and energy currents, see panel (b) for currents flowing in the same direction and panel (b) 
for currents flowing in opposite directions.

\begin{figure}
\begin{centering}
\includegraphics[width=13cm]{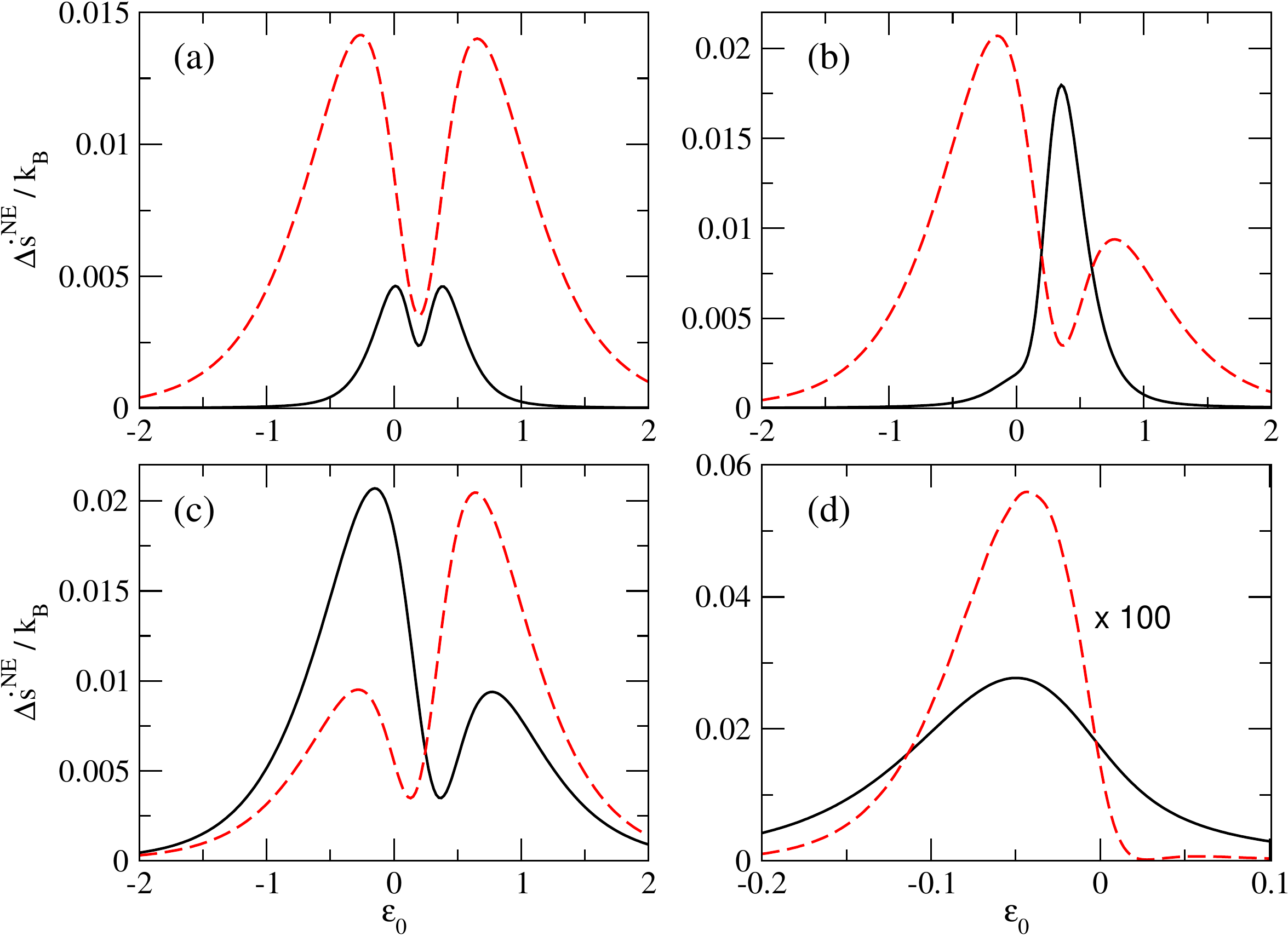}
\end{centering}
\caption{NE entropy production rate $\Delta \dot{S}^{\rm NE}$ versus the energy level $\varepsilon_0$
for different transport regimes. 
$\Delta \dot{S}^{\rm NE}$ is always a positive quantity.
(a) Only temperature differences ($\mu_L=\mu_R=0.2$) $k_B T_L=0.1$, $k_B T_R=0.05$ (solid line)
and $k_B T_L=0.1$, $k_B T_R=0.3$ (dashed line).
(b) Both chemical potential and temperature differences ($\mu_L=0.3$, $\mu_R=0.2$)
 $k_B T_L=0.1$, $k_B T_R=0.05$ (solid line)
and $k_B T_L=0.1$, $k_B T_R=0.3$ (dashed line).
(c) Both temperature and chemical potential differences ($k_B T_L=0.1$, $k_B T_R=0.3$)  
$\mu_L=0.3$, $\mu_R=0.2$ (solid line)
and $\mu_L=0.2$, $\mu_R=0.3$ (dashed line). 
(d) Comparison with results of Fig. 3(b) in Ref.~\cite{Esposito:2015}.
($\mu_L=0.05$, $\mu_R=0.0$) $k_B T_L=0.026$, $k_B T_R=26/30.10^{-3}$, 
strong coupling $v_L=v_R=0.2$ (solid line) and weak coupling $v_L=v_R=0.02$ (dashed line, amplitude
rescaled by a factor $\times 100$). 
The other parameters are $t_L=t_R=2.0$ and $v_L=v_R=0.25$ (when not specified otherwise). 
All parameters are given in dimension of energy in [eV].}
\end{figure}   
\label{fig:Sentropy_rate_vs_e0}

In panel (d) Fig. 1, we have tried to reproduce the results shown in Fig. 3(b) of 
\cite{Esposito:2015}. 
The results are qualitatively reproduced apart from the behaviour of the over whole 
amplitude of the entropy production rate.
Indeed, in our model, the transmission coefficient $T(\omega)$ has roughly a Lorentzian
lineshape, with a maximum amplitude of unity (whatever the values of the parameters are) and
a width which scales approximately as $\sum_\alpha v_\alpha^2/t_\alpha$ versus the 
coupling parameters between the central region $C$ and the reservoirs. 
Hence the width of $T(\omega)$ increases with the strength of the coupling to the reservoirs, 
and therefore the currents will always have a larger values when increasing
the strength of this coupling.
Consequently, the entropy production rate $\Delta \dot{S}^{\rm NE}$ defined by 
Eq.(\ref{eq:NESentropy_rate}) is always larger for larger values of the coupling
to the reservoirs.

In Figure 2, we show how the NE entropy production rate $\Delta \dot{S}^{\rm NE}$
depends on the NE conditions, i.e. on the chemical potential difference $\Delta\mu$ 
(see left panels (a) and (c) in Fig.2) or on temperature differences $\Delta T$ 
between the reservoirs (see right panels (b) and (d) in Fig.2).
First it is important to note that, for all the parameters used, the NE entropy production 
rate $\Delta \dot{S}^{\rm NE}$ is always a positive quantity (as expected).
Furthermore, $\Delta \dot{S}^{\rm NE}$ increases when the NE conditions are more important, 
i.e. when $\Delta\mu$ or $\Delta T$ increases. 
In other words, the more the system is out of equilibrium, the larger the entropy production 
becomes.

\begin{figure}
\centering
\includegraphics[width=13cm]{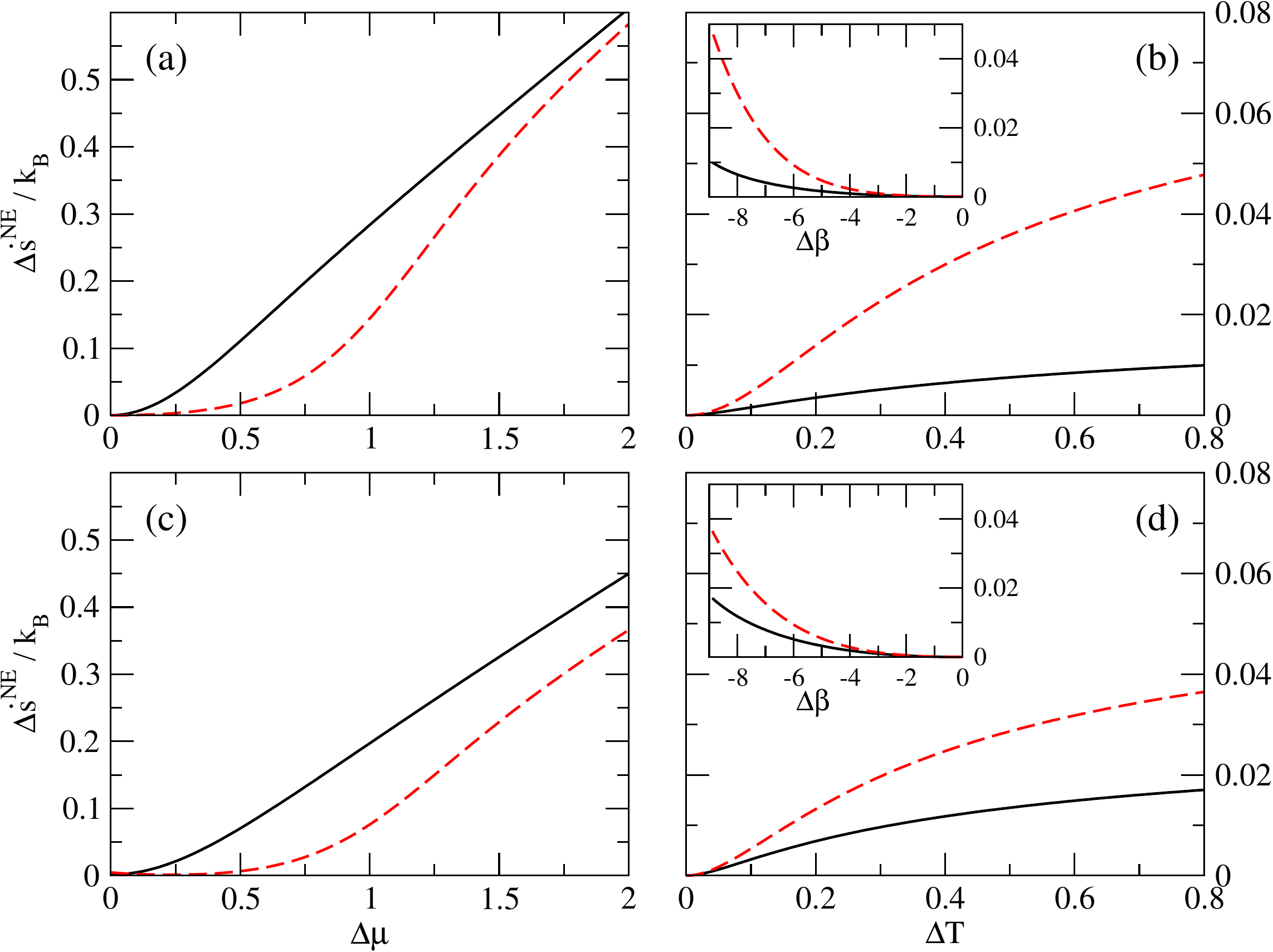}
\caption{NE entropy production rate $\Delta \dot{S}^{\rm NE}$ versus the
temperature difference ($\Delta T = T_L - T_R$ and $\Delta\beta=\beta_L-\beta_R$) 
and/or the chemical potential 
difference ($\Delta\mu=\mu_L-\mu_R$).
$\Delta \dot{S}^{\rm NE}$ is always a positive quantity, and increases when
$\Delta\mu$ or $\Delta T$ ($\vert\Delta\beta\vert$) increases.
The solid (dashed) lines are for the resonant (off-resonant) transport regime, i.e.
$\varepsilon_0 \sim (>)\ \mu^{\rm eq}$ when $\Delta\mu=0$.
(a) System at a unique temperature $k_B T_L=k_B T_R=0.1$.
(b) System with a unique chemical potential $\mu_L=\mu_R=0.2$. In the inset,
we also show the dependence of $\Delta \dot{S}^{\rm NE}$ vs $\Delta\beta$.
(c) System with a temperature difference $k_B T_L=0.1, k_B T_R=0.2$.
(d) System with a chemical potential difference $\mu_L=0.35, \mu_R=0.05$.
The inset shows the $\Delta \dot{S}^{\rm NE}$ vs $\Delta\beta$.
The other parameters are $t_L=t_R=2.0$ and $v_L=v_R=0.25$ (given in [eV]).
For the dependence on $\Delta\mu$, we take
$\mu_L=\mu^{\rm eq}+\Delta\mu/2$ and $\mu_R=\mu^{\rm eq}-\Delta\mu/2$
with $\mu^{\rm eq}=0.2$.
For the dependence on $\Delta T$, we take
and $T_R=T^0=0.1$, $T_L=T^0+\Delta T$ (hence $\Delta\beta < 0$ for $\Delta T >0$).}
\end{figure}

One should note that the dependence of $\Delta \dot{S}^{\rm NE}$ on $\Delta\mu$ shows some
form of linearity when $\Delta\mu \ge \Delta\mu^*$. This is simply due to the fact that
the currents saturate: $\langle j_{Q,E}(\Delta\mu) \rangle = I^{\rm sat}_{Q,E}$ when  
$\Delta\mu \ge \Delta\mu^*$. Indeed, in the saturation region, increasing $\Delta\mu$
does not change the value of the moments $M_n$
as the transmission $T(\omega)$ is zero in energy range where $f_L(\omega)-f_R(\omega)$
is modified by an increase of $\Delta\mu$.
In this regime, one can easily see that the dependence of $\Delta \dot{S}^{\rm NE}$ on $\Delta\mu$
is simply linear with a slope given by $(\beta_L+\beta_R) I^{\rm sat}_Q / 2$.
Furthermore, the slope is maximal when $\beta_L=\beta_R$ and smaller for any $\beta_L \ne \beta_R$
as clearly exemplified by the results shown in panels (a) and (c) in Figure 2.
Such a saturation regime does not exist for increase  $\Delta T$ differences (at fixed $\Delta\mu$)
as shown in the panels (b) and (d) in Figure 2.

\section{Nonequilibrium Gibbs~-~von Neumann entropies}
\label{sec:entropy_more}

In the previous Section, we have shown how the NE entropy production
rate is related to the charge and energy currents.
We have also shown that the NE steady state can be considered as a pseudo
equilibrium state with a corresponding (time-independent) density matrix 
which is given in the form of a generalised Gibbs ensemble.
It is therefore be very interesting to be able to define a NE entropy \cite{Penrose:1979,Maroney:2007}
from 
the NE density matrix by using the equivalence between pseudo equilibrium states 
and equilibrium states.
In other words: when we build a NE entropy from the equilibrium expression \cite{Penrose:1979,Maroney:2007} 
$S= k_B \langle\ln\mathcal{W}\rangle=-k_B {\rm Tr}[\rho\ln\rho]$,
which density matrix should be used?

\subsection{Which density matrix?}
\label{sec:gibbs_entropy}

The first natural choice would be to take the NE density matrix $\rho^{\rm NE}$ 
derived in the previous Section.
However such a choice does not bring any information about entropy production under 
the NE conditions.
Indeed, if we consider the asymptotic operator $A^{(+)}$ being
obtained from a unitary transformation ($\Omega^{(+)}\Omega^{(+)-1}= \mathbbm{1}$), 
we can show that for any function $f(A)$:
\begin{equation}
\label{eq:TrA+B+}
\begin{split}
{\rm Tr}[A^{(+)}f(A^{(+)})]=
{\rm Tr}[A^{(+)}\Omega^{(+)}f(A)\Omega^{(+)-1}]=
{\rm Tr}[\Omega^{(+)}Af(A)\Omega^{(+)-1}]={\rm Tr}[Af(A)] \ .
\end{split}
\end{equation}
By taking $A=\rho_0$ and $f(.)=\ln(.)$, one easily see that
${\rm Tr}[\rho^{\rm NE}\ln\rho^{\rm NE}]={\rm Tr}[\rho_0\ln\rho_0]$.
The quantity ${\rm Tr}[\rho_0\ln\rho_0]=\sum_{\alpha={L,C,R}} {\rm Tr}_{(\alpha)}[\rho_\alpha\ln\rho_\alpha]$ 
defines the entropy of the three separated $L,C,R$ regions.
It does not contains any information about the charge and energy currents following
through the entire system under general NE conditions.

Another possibility would be to take the NE average of the
density matrix of the coupled system at equilibrium, i.e.
$S^{\rm NE} = -k_B {\rm Tr}[\rho^{\rm NE}\ln\rho^{\rm eq}]$ where
$\rho^{\rm eq}=Z^{-1} {\rm exp} \left\{ -\beta^{\rm eq}(H-\mu^{\rm eq}N) \right\}$.
However, from the intertwining relation, we have
$\Omega^{(+)-1}H\Omega^{(+)}=H_0$ and $\Omega^{(+)-1}N\Omega^{(+)}=N$ \cite{note:N+},
and
we obtain 
$S^{\rm NE} = k_B {\rm Tr}[\rho^{\rm NE}\ln\rho^{\rm eq}] = -k_B {\rm Tr}[\rho_0\ln\rho_0^{\rm eq}]$
with $\rho^{\rm eq} \propto {\rm exp} \left\{ -\beta^{\rm eq}(H_0-\mu^{\rm eq}N) \right\}$.
Such an entropy contains some information about the NE conditions, 
considering that $\beta_{L,R}$ and $\mu_{L,R}$ in $\rho_{L,R}$ are different
from the equilibrium $\beta^{\rm eq}$ and $\mu^{\rm eq}$.
However this entropy is defined from the non-interacting Hamiltonian only, 
and it lacks the presence of the operator $W$ which is the generator of the different 
charge and energy currents. 
Hence such an entropy does not contain any information about the fluxes which are the 
responses to the applied forces $\Delta\mu$ and $\Delta T$.  

One has to go back to the definition of the NE steady state
averages given in Eq.~(\ref{eq:NEaverage}). 
The asymptotic time-dependence, in such average, has been passed on to the NE density 
matrix which we use to calculate the average of quantum operators.
Hence, it follows that one should define the entropy from the NE average
of the nominal density matrix $\rho_0$, i.e.
$S^{\rm NE} = -k_B {\rm Tr}[\rho^{\rm NE}\ln\rho_0]$.

As the density matrix $\rho_0$ is the direct product of the individual
density matrices of each separate $L,C,R$ regions, it is easy to show that
\begin{equation}
\label{eq:NEgibbs_entropy}
S^{\rm NE} = -k_B {\rm Tr}[\rho^{\rm NE}\ln\rho_0] = S_L^{\rm NE} + S_C^{\rm NE} + S_R^{\rm NE}
\quad {\rm where} \quad
S_\alpha^{\rm NE} = -k_B {\rm Tr}_{(\alpha)}[\rho^{\rm NE}_{{\rm red},\alpha}\ln\rho_\alpha] \ .
\end{equation}
$S_\alpha^{\rm NE}$ is the contribution of the region $\alpha$ 
and $\rho^{\rm NE}_{{\rm red},\alpha}$ is the corresponding reduced density matrix
obtained from $\rho^{\rm NE}_{{\rm red},\alpha}={\rm Tr}_{(\beta,\beta')}[\rho^{\rm NE}]$
with $\beta,\beta'=L,C,R$ and $\beta\ne\beta'\ne\alpha$.
For example, the NE reduced density matrix in the central region $\rho^{\rm NE}_{{\rm red},C}$ 
is obtained from $\rho^{\rm NE}_{{\rm red},\alpha}={\rm Tr}_{(L,R)}[\rho^{\rm NE}]$.

The corresponding entropy $S_C^{\rm NE}$ has been the object of recent studies
\cite{Ludovico:2014,Esposito:2015,Topp:2015,Bruch:2016,Solano:2016} but it is clearly
only a part of the entire entropy production in the system.
For example, the contributions $S_L^{\rm NE}$ and $S_R^{\rm NE}$ are different from their
(isolated) equilibrium counter parts 
$S_\alpha^{\rm eq}= -k_B {\rm Tr}_{(\alpha)}[\rho_\alpha\ln\rho_\alpha]$
($\alpha=L,R$) since  $\rho^{\rm NE}_{{\rm red},\alpha}\ne\rho_\alpha$.

We now further comment on this point. For that, we consider small deviations from the 
equilibrium where
$\mu_{L,R}=\mu\pm\frac{1}{2}\Delta\mu$ and
$\beta_{L,R}=\beta\pm\frac{1}{2}\Delta\beta$ with
$\Delta\mu\ll1$ (hence $\Delta_\mu\ll1$) and $\Delta\beta\ll1$. Hence
\begin{equation}
\label{eq:rho_NE_entropy}
\begin{split}
\rho^{\rm NE} & \approx Z^{-1}\ {\rm exp} \left\{-\beta ( H_{L+R}^{(+)} - \mu N_{L+R}^{(+)} ) \right\}
		\left( 1 + \Delta_\mu Q^{(+)} \right) \left( 1 - \Delta\beta E^{(+)} \right)
     \ \rho_C(H_C^{(+)}) \ \\
               & \approx Z^{-1}\ {\rm exp} \left\{-\beta ( H_{L+R}^{(+)} - \mu N_{L+R}^{(+)} ) \right\} \ \rho_C(H_C^{(+)})
		\left( 1 + \Delta_\mu Q^{(+)} - \Delta\beta E^{(+)} \right) \ ,
\end{split}
\end{equation}
where we kept only the lowest order terms in $\Delta_\mu$ and $\Delta\beta$.
Furthermore, if we assume a lowest order expansion of the density matrices
$e^{-\beta ( H_\alpha^{(+)} - \mu N_\alpha^{(+)} )} \approx e^{-\beta ( H_\alpha - \mu N_\alpha )}$
and $\rho_C(H_C^{(+)})\approx \rho_C(H_C)$, one gets:
\begin{equation}
\label{eq:rho_NE_entropy_2}
\rho^{\rm NE}_{{\rm red},L}={\rm Tr}_{(C,R)}[\rho^{\rm NE}]
\approx 
\rho_L {\rm Tr}_{(C,R)}\left[ 1 + \Delta_\mu Q^{(+)} - \Delta\beta E^{(+)} \right] \ ,
\end{equation}
and therefore
\begin{equation}
\label{eq:NE_entropy_L}
\begin{split}
S_L^{\rm NE} & = -k_B {\rm Tr}_{(L)}\left[\rho^{\rm NE}_{{\rm red},L}\ln\rho_L\right] \\
& \approx -k_B {\rm Tr}_{(L)}\left[\rho_L\ln\rho_L\right] 
- k_B {\rm Tr}_{(L)}\left[\rho_L {\rm Tr}_{(C,R)}\left[\Delta_\mu Q^{(+)} - \Delta\beta E^{(+)} \right]\ln\rho_L \right] \ .
\end{split}
\end{equation}
The first term in the above equation is simply the entropy $S^{\rm eq}_L$ of the isolated $L$ region 
with the associated grand canonical density matrix given by Eq.~(\ref{eq:rho_LR}).
The second term can be re-arranged as follows:
$- k_B {\rm Tr}_{(L)} [\rho_L {\rm Tr}_{(C,R)} [\Delta_\mu Q^{(+)} - \Delta\beta E^{(+)} ] \ln\rho_L ] 
=- k_B {\rm Tr}[ \rho_L (\Delta_\mu Q^{(+)} - \Delta\beta E^{(+)} ) \ln\rho_L ]
=  {\rm Tr}[ (\Delta_\mu Q^{(+)} - \Delta\beta E^{(+)} ) ( -k_B \rho_L \ln\rho_L ) ] $.
Finally, we have
\begin{equation}
\label{eq:NE_entropy_Lbis}
S_L^{\rm NE} \approx S^{\rm eq}_L + {\rm Tr}\left[\frac{\Delta\mathcal{S}^{\rm NE}}{k_b} \mathcal{S}^{\rm eq}_L \right] \ ,
\end{equation}
where $\Delta\mathcal{S}^{\rm NE}$ is the operator defining the entropy production in Eq.~(\ref{eq:NESentropy}), 
and $\mathcal{S}^{\rm eq}_L = -k_B \rho_L\ln\rho_L$
with $S^{\rm eq}_L = {\rm Tr}_{(L)}[\mathcal{S}^{\rm eq}_L]$. 
Similar expressions can be found for $S_C^{\rm NE}$ and $S_R^{\rm NE}$. 

The result show that, under general NE conditions, NE entropy is produced in the central region and in
the reservoirs as well.  Such an entropy is always related to the charge and energy
currents flowing at the interfaces between the central region and the $L$ and $R$ regions.

The full calculation of the entropy from Eq.~(\ref{eq:NEgibbs_entropy}) is a non trivial task, especially
for arbitrary interaction $V_C^{\rm int}$ in the central region. This can however achieved
by either determining the asymptotic scattering states $\vert L_k^{(+)}\rangle = \Omega^{(+)}\vert L_k\rangle$
for the $L$ region (and for the states $\vert R_k\rangle$ and $\vert C_n\rangle$
for the $R$ and $C$ regions respectively).
Following \cite{Han:2006,Han:2007,Han:2007b,Han:2010,Han:2010b,Topp:2015}, the scattering states of the $L$ region,
for the model described in Section \ref{sec:example_DeltaS}, are given by:
$\vert L_k^{(+)}\rangle = \vert L_k\rangle + v_L G^r(\epsilon_{Lk})\vert C_n\rangle
+ \sum_{\alpha=L,R; k'} v_L v_{\alpha} G^r(\epsilon_{Lk}) / ( \epsilon_{Lk} - \epsilon_{\alpha k'} + i0^+ ) \vert\alpha_{k'}\rangle
$.
For the non-interacting case in Section \ref{sec:example_DeltaS}, the calculations of the entropy can also
be easily performed using the NEGF formalism which we consider in the next Section.

\subsection{An example for the entropy of the central region}
\label{sec:example_SGibbs}

We now consider numerical calculations for the Gibbs~-~von Neumann entropy using the
single level model described in Section \ref{sec:example_DeltaS}.
We have shown that the NE steady state can be considered as a pseudo
equilibrium state with a corresponding generalised Gibbs ensemble given
by $\rho^{\rm NE}$.
Following the same principles of equilibrium statistical mechanics, one 
can defined from the generalised Gibbs ensemble a local NE distribution
functions \cite{Ness:2014a} in the $L,C,R$ regions.
From these NE distributions functions, one can also defined the corresponding
Gibbs~-~von Neumann entropies.
For example, the NE entropy $S_C^{\rm NE}$ in the central region $C$ 
\HN{can be defined as follows \cite{Note:comments_on_entropy}:}
\begin{equation}
\label{eq:NE_S_C}
S_C^{\rm NE}(\Delta\mu,\Delta T) 
= - k_B \int \frac{{\rm d}\omega}{2\pi}\ A_C(\omega) 
\left[ f_C^{\rm NE}(\omega) \ln f_C^{\rm NE} + (1-f_C^{\rm NE}(\omega)) \ln (1-f_C^{\rm NE}) 
\right] \ ,
\end{equation}
where $f_C^{\rm NE}(\omega)$ is the NE distribution function of the central region
and $A_C(\omega)$ is the corresponding spectral function defined from the NEGF
as  
$A_C(\omega) = - {\rm Im} G^r(\omega) / \pi$.
It should be noted that the entropy $S_C^{\rm NE}$ is only a part of the total entropy 
$S^{\rm NE}$ in Eq.(\ref{eq:NEgibbs_entropy}), 
which is produced in the entire system under the general NE conditions.

For non-interacting system 
\HN{considered in Sec.~\ref{sec:example_DeltaS}}, 
the NE distribution function in the central region
is just an weighted averaged of the equilibrium Fermi distributions of the reservoirs 
$f_C^{\rm NE}(\omega) = [ \Gamma_L(\omega) f_L(\omega) + \Gamma_R(\omega) f_R(\omega) ] / \Gamma_{L+R}(\omega)$.

\begin{figure}
\label{fig:NEentropy_Gibbs_vs_e0}
\centering
\includegraphics[width=13cm]{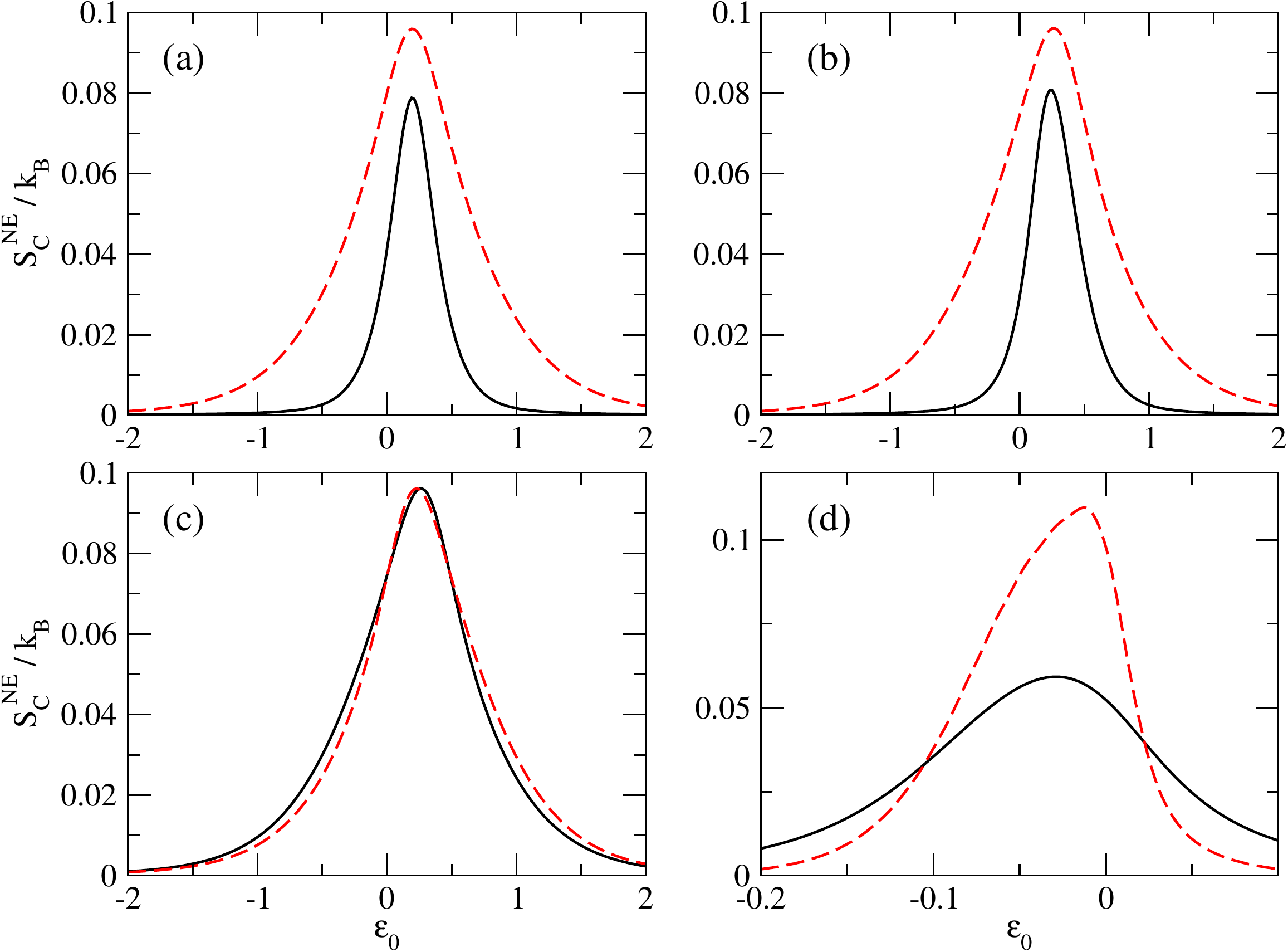}
\caption{Gibbs~-~von Neumann NE entropy for the central region $S_C^{\rm NE}$ versus 
the energy level $\varepsilon_0$
for the different transport regimes considered in Figure 1. 
The Gibbs NE entropy $S_C^{\rm NE}$ is always a positive quantity as expected.
(a) ($\mu_L=\mu_R=0.2$) $k_B T_L=0.1$, $k_B T_R=0.05$ (solid line)
and $k_B T_L=0.1$, $k_B T_R=0.3$ (dashed line).
(b) Both chemical potential and temperature differences ($\mu_L=0.3$, $\mu_R=0.2$)
 $k_B T_L=0.1$, $k_B T_R=0.05$ (solid line)
and $k_B T_L=0.1$, $k_B T_R=0.3$ (dashed line).
(c) Both temperature and chemical potential differences ($k_B T_L=0.1$, $k_B T_R=0.3$)  
$\mu_L=0.3$, $\mu_R=0.2$ (solid line)
and $\mu_L=0.2$, $\mu_R=0.3$ (dashed line). 
(d) Comparison with results of Fig. 3(b) in Ref.~\cite{Esposito:2015}.
($\mu_L=0.05$, $\mu_R=0.0$) $k_B T_L=0.026$, $k_B T_R=26/30.10^{-3}$, 
strong coupling $v_L=v_R=0.2$ (solid line) and weak coupling $v_L=v_R=0.02$ (da$S_C^{\rm NE}$shed line).
The other parameters are $t_L=t_R=2.0$ and $v_L=v_R=0.25$ (when not specified otherwise) 
and given in [eV].}
\end{figure}   

Figure 3 shows the dependence of the entropy $S_C^{\rm NE}$ calculated
for different transport regime. Once more, we can see that $S_C^{\rm NE}$ is always a positive
quantity. The positiveness of $S_C^{\rm NE}$ is obtained when the system has a single chemical 
potential (see panel (a) in Fig. 3) as well as when there are both a chemical potential and
temperature differences between the reservoirs (see panels (b) and (c) in Fig. 3).
In panel Fig. 3.(d), we show the behaviour of the entropy for the same parameter used in Fig. 1.(d)
and we recover the same qualitative behaviour as shown in Fig. 3(b) of \cite{Esposito:2015}. 
The amplitude of the entropy is larger in the weak coupling limit in comparison to the strong 
coupling limit to the reservoirs.
One should however note that in Fig. 3, $S_C^{\rm NE}$ has the dimension of an entropy, i.e.
[energy]/[temperature], while in Fig. 3(b) of \cite{Esposito:2015} and Fig. 1, we are dealing 
with an entropy production rate, 
i.e. a quantity with dimension [energy]/[temperature $\times$ time].

\begin{figure}
\centering
\includegraphics[width=13cm]{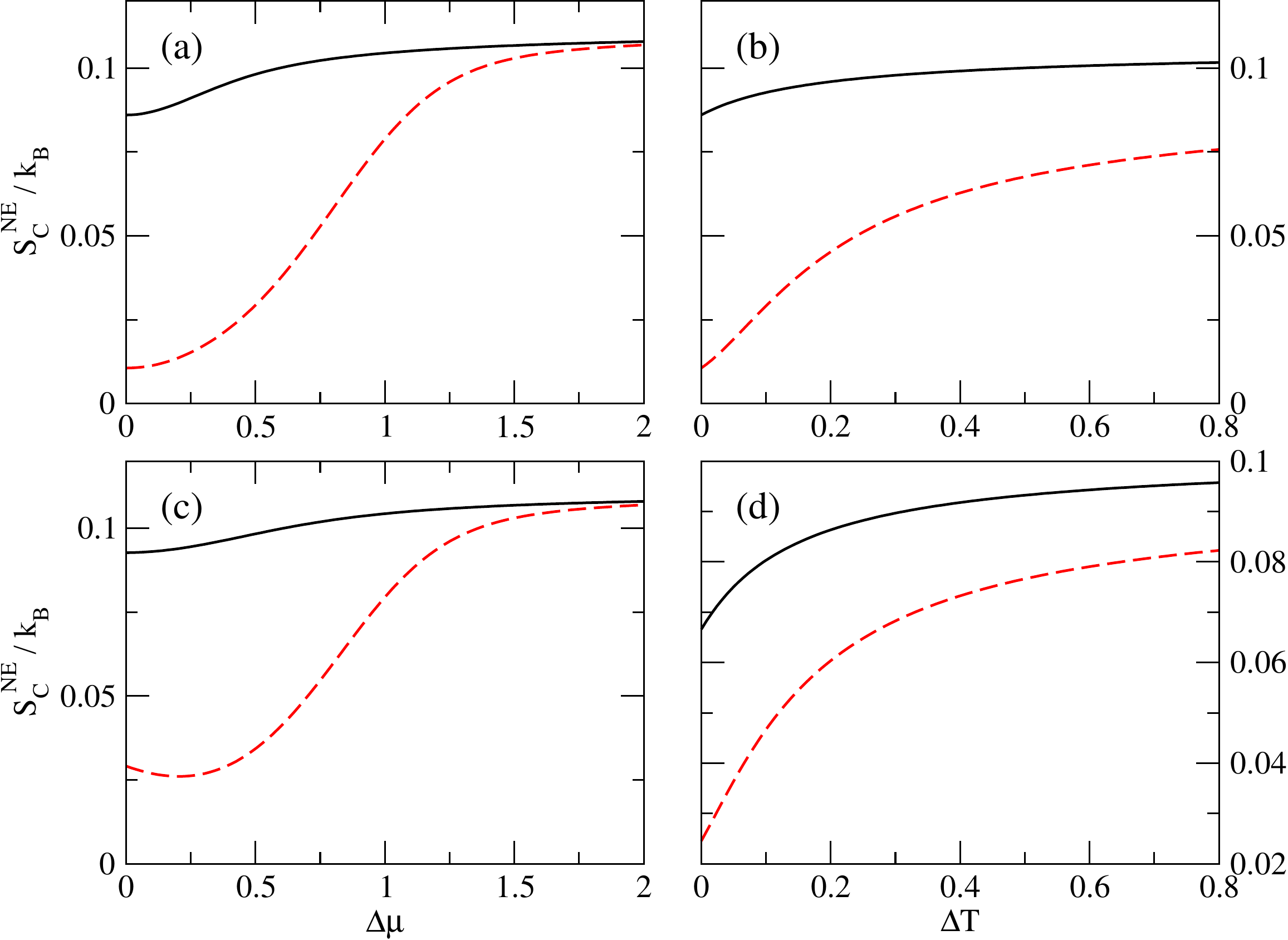}
\caption{Gibbs~-~von Neumann NE entropy for the central region  $S_C^{\rm NE}$ versus the
temperature difference ($\Delta T = T_L - T_R$ and $\Delta\beta=\beta_L-\beta_R$) 
and/or the chemical potential 
difference ($\Delta\mu=\mu_L-\mu_R$).
$\Delta \dot{S}^{\rm NE}$ is always a positive quantity, and increases when
$\Delta\mu$ or $\Delta T$ ($\vert\Delta\beta\vert$) increases.
The solid (dashed) lines are for the resonant (off-resonant) transport regime, i.e.
$\varepsilon_0 \sim (>)\ \mu^{\rm eq}$ when $\Delta\mu=0$.
(a) System at a unique temperature $k_B T_L=k_B T_R=0.1$.
(b) System with a unique chemical potential $\mu_L=\mu_R=0.2$. In the inset,
we also show the dependence of $\Delta \dot{S}^{\rm NE}$ vs $\Delta\beta$.
(c) System with a temperature difference $k_B T_L=0.1, k_B T_R=0.2$.
(d) System with a chemical potential difference $\mu_L=0.35, \mu_R=0.05$.
The inset shows the $\Delta \dot{S}^{\rm NE}$ vs $\Delta\beta$.
The other parameters are $t_L=t_R=2.0$ and $v_L=v_R=0.25$ (given in [eV]).
For the dependence on $\Delta\mu$, we take
$\mu_L=\mu^{\rm eq}+\Delta\mu/2$ and $\mu_R=\mu^{\rm eq}-\Delta\mu/2$
with $\mu^{\rm eq}=0.2$.
For the dependence on $\Delta T$, we take
and $T_R=T^0=0.1$, $T_L=T^0+\Delta T$ (hence $\Delta\beta < 0$ for $\Delta T >0$).}
\end{figure}

In Figure 4, we show the dependence of the entropy $S_C^{\rm NE}$
on the NE conditions, i.e. on  the chemical potential difference $\Delta\mu$, as shown
in the left panels (a) and (c), and on temperature differences $\Delta T$ between the reservoirs, 
as shown in the right panels (b) and (d).
One can see that, for the range of parameters we used, the NE entropy production  $S_C^{\rm NE}$
is once more a positive quantity (as expected).
Furthermore, the entropy $S_C^{\rm NE}$ increases with the NE conditions, i.e. it increases
for increasing values of $\Delta\mu$ and/or $\Delta T$.
We also observe a saturation regime in $S_C^{\rm NE}$ with increasing $\Delta\mu$.
In the saturation regime, an increase of $\Delta\mu$ changes the features of the NE distribution
function $f_C^{\rm NE}(\omega)$ in an energy range where the spectral function $A_C(\omega)$
has no weight, i.e. where $A_C(\omega)=0$. Therefore the energy integral in Eq.~(\ref{eq:NE_S_C})
does not change with increasing $\Delta\mu$ and the entropy $S_C^{\rm NE}$ saturates.

\HN{
Finally, one should note that the calculation of the Gibbs~-~von Neumann NE entropy 
for the central region as well as the calculation of the entropy production rate
Eq.~(\ref{eq:NESentropy_rate}) can also be performed when interactions are present 
in the central region. Our expressions for the entropies are generally applicable
to the cases with and without interactions in the central region.
The latter is directly related to the charge and energy currents which can be 
calculated for different kinds of interaction in the central region. For example,
in \cite{Dash:2010,Ness:2010,Dash:2011,Dash:2012,Ness:2012}, we have studied the effect
of electron-vibration interaction on the electron current.
For Gibbs~-~von Neumann NE entropy, one can also defined a NE distribution
function $f_C^{\rm NE}$ which contains all the effects of the interactions as 
shown in \cite{Ness:2014a}.
The interactions will affect the entropy production, however we expect that,
with the so-called conservative approximations for the interaction, the positiveness 
of the entropy will be conserved. In the presence of interaction with extra degrees 
of the freedom (vibration or other boson modes), the contribution of their respective
entropy production will need to be taken into accout.
However such an in-depth study is out of the scope of the present paper.
}

%%%%%%%%%%%%%%%%%%%%%%%%%%%%%%%%%%%%%%%%%%
\section{Discussion and conclusion}
\label{sec:ccl}

We have studied the steady state NE thermodynamical properties
of an open quantum system connected to two reservoirs $\alpha$, the latter are acting 
as equilibrium (particle and heat) baths with their respective temperature $T_\alpha$ and 
chemical potential $\mu_\alpha$.
We have shown that the steady state of the entire system can be seen as a
pseudo equilibrium state. The corresponding NE density matrix is expressed
in the form of a generalised Gibbs ensemble 
$\rho^{\rm NE}=e^{-\bar\beta(H-\bar\mu N + \sum_{a=Q,E} \lambda_a \mathcal{J}_a )}/Z$. 

The NE density matrix is time independent and built from the so-called
conserved quantities: the total Hamiltonian $H$ and the total number
of electrons $N$ and the $\mathcal{J}_{Q,E}$ quantities which are related to the fluxes
of charge and energy flowing in between the central region $C$
and the reservoirs.
We have given different forms for the NE density matrix and shown their mutual
equivalence. 
The extra terms entering the definition of $\rho^{\rm NE}$ which do not exist in the
equilibrium grand canonical representation have been clearly identified and have been 
shown to be related to the entropy production in the entire system.
From their expression, the entropy production rate is given in terms of the charge 
and energy currents.

We have calculated such an entropy production rate for a model system consisting in
a single electron resonance coupled to two Fermi reservoirs. 
Numerical results performed for different transport regimes have shown that the 
entropy production rate is always a positive quantity.

Furthermore, based upon the pseudo equilibrium properties of the steady state,
we have also calculated a Gibbs~-~von Neumann entropy for the entire system. 
Our results show that the NE conditions create extra entropy in the central region 
as well as in the reservoirs. 
The former can be derived from the equilibrium expression of the entropy by using 
the appropriate NE distribution function in the central region.

Our numerical results for the entropy production and production rate corroborate 
and expand earlier 
studies \cite{Ludovico:2014,Esposito:2015,Topp:2015,Bruch:2016,Solano:2016}.
These results also open a new route for determining the NE thermodynamical properties 
of quantum open systems under general conditions. 
For example the corresponding NE specific heat or charge susceptibility \cite{Ness:2012} 
can be directly obtained from the derivative of the entropy versus the applied temperature 
or chemical potential biases.

%%%%%%%%%%%%%%%%%%%%%%%%%%%%%%%%%%%%%%%%%%
\vspace{6pt} 

%%%%%%%%%%%%%%%%%%%%%%%%%%%%%%%%%%%%%%%%%%
%% optional
%\supplementary{}

%%%%%%%%%%%%%%%%%%%%%%%%%%%%%%%%%%%%%%%%%%
\acknowledgments{The author thanks Benjamin Doyon and Lev Kantorovich 
for fruitful discussions.
\HN{The UK EPSRC is acknowledged for financial support under Grant No. EP/J019259/1. }
}

%%%%%%%%%%%%%%%%%%%%%%%%%%%%%%%%%%%%%%%%%%

\bibliographystyle{mdpi}

\begin{thebibliography}{-------}
\bibitem{note:N+}
In Appendix A of \cite{Ness:2014c}, we have demonstrated the
intertwining relation $H\Omega^{(+)}=\Omega^{(+)}H_0$. Following the same
steps, we can also show that $N\Omega^{(+)}=\Omega^{(+)}N$ when one assumes
that the total number of electron operator $N$ commutes with the Hamiltonian
$H$. Hence we have not only $[H_\alpha, N_\beta] = 0$ but also
$[H, \sum_\beta N_\beta] = 0$. In other words, the total number of electron is
conserved and $[W, N_L + N_C + N_R] = 0$, implying to all the electrons coming
out (in) the reservoirs are going in (out) the central region $C$.
\end{thebibliography}

%=====================================
% References, variant A: internal bibliography
%=====================================

\renewcommand\bibname{References}

\end{document}